\documentclass[twocolumn, amssymb,amsmath,pra,superscriptaddress]{revtex4-1}
\usepackage{graphicx}
\usepackage{float}
\usepackage{ulem}

\begin{document}
\title{From membrane-in-the-middle to mirror-in-the-middle with a high-reflectivity sub-wavelength grating}
\author{Corey Stambaugh}
\affiliation{National Institute of Standards and Technology , Gaithersburg MD, 20899, USA}
\author{Haitan Xu}
\affiliation{Joint Quantum Institute, University of Maryland, College Park, MD 20742, USA}
\affiliation{National Institute of Standards and Technology , Gaithersburg MD, 20899, USA}
\author{Utku Kemiktarak}
\affiliation{National Institute of Standards and Technology , Gaithersburg MD, 20899, USA}
\affiliation{Joint Quantum Institute, University of Maryland, College Park, MD 20742, USA}
\author{Jacob Taylor}
\affiliation{National Institute of Standards and Technology , Gaithersburg MD, 20899, USA}
\affiliation{Joint Quantum Institute, University of Maryland, College Park, MD 20742, USA}
\author{John Lawall}
\email{lawall@nist.gov}
\affiliation{National Institute of Standards and Technology , Gaithersburg MD, 20899, USA}

\begin{abstract}
We demonstrate a ``membrane in the middle'' optomechanical system using a silicon nitride membrane patterned as a subwavelength grating.  The grating has a reflectivity of over 99.8$\%$,  effectively creating two sub-cavities, with free spectral ranges of 6~GHz, optically coupled via photon tunneling.  Measurements of the transmission and reflection spectra show an avoided crossing where the two sub-cavities simultaneously come into resonance, with a frequency splitting of 54~MHz.  We derive expressions for the lineshapes of the symmetric and antisymmetric modes at the avoided crossing, and infer the grating reflection, transmission, absorption, and scattering through comparison with the experimental data.  
\end{abstract}
\maketitle

\section{Introduction}
Dramatic improvements in optical and mechanical design and fabrication enable new regimes of light-matter interactions~\cite{Aspelmeyer2013}, including observation of long-predicted effects such as cooling to the ground state of a mechanical oscillator~\cite{OConnell2010}, pondermotive squeezing~\cite{Brooks2012,Safavi-Naeini2013,purdy_strong_2013}, and new ultra-sensitive sensors. Of the different approaches,
the ``membrane in the middle'' platform for cavity optomechanics, discussed~\cite{bhattacharya_optomechanical_2008} and first demonstrated~\cite{Thompson2008a} in 2008, is notable  
for the fact that it decouples the technical demands on the optical and mechanical resonators.
It has been used by a number of groups worldwide, for example to 
observe radiation pressure shot noise~\cite{purdy_observation_2013}, demonstrate optomechanical transparency~\cite{karuza_optomechanically_2013},
generate squeezed light~\cite{purdy_strong_2013}, and optically hybridize distinct mechanical modes~\cite{shkarin_optically_2014}.  
In addition, a number of recent theoretical proposals exploit the possibilities inherent in a ``membrane in the middle'' system
with membranes of high reflectivity, including achieving very high optomechanical coupling to the collective modes of
an array of such membranes~\cite{xuereb_collectively_2013,Xuereb2012}, creating interference between adjacent longitudinal cavity modes~\cite{genes_enhanced_2013}, new approaches to force sensing with sensitivity exceeding the standard quantum limit~\cite{Xu2013a},
and studies of quantum nonlinear optics~\cite{Xu2013b}.  However, to date, experimental efforts have focused on low-reflectivity membranes, where the perturbations of the cavity modes, though not their frequencies, follow adiabatically with the mechanical motion~\cite{Thompson2008a,sankey_strong_2010,flowers-jacobs_fiber-cavity-based_2012}.  In contrast, at high reflectivity, the membrane-mirror effectively divides the cavity into left and right modes~\cite{xuereb_dynamical_2012}, leading to the potential for non-adiabatic corrections to the mode structure with the motion of the membrane and a variety of novel optomechanical phenomena~\cite{Rabl2011,Komar2013,Ludwig2012,Xu2013a,Xu2013b}.

Obtaining high reflectivity with a uniform dielectric membrane is not possible, however, as one is limited by the Fresnel equations
governing reflection at dielectric interfaces as
applied to actual materials.  High membrane reflectivity can, however, be achieved if the membrane is not uniform but patterned as a
photonic crystal structure.  It has long been recognized that the reflectivity of such a structure
can approach unity at normal incidence~\cite{fan_analysis_2002}.  
Motivated by these ideas, mechanically compliant photonic crystal structures have been developed in InP~\cite{antoni_deformable_2011,antoni_nonlinear_2012}
and silicon nitride~\cite{BuiPhotXtalAPL2012}; in the latter case, importantly, little or no degradation of the mechanical quality factor was observed.
  A variant on this approach is provided by a sub-wavelength diffraction grating, in which only zero-order diffraction is allowed
and very high reflectivity can be engineered.  Such ``high-contrast grating'' (HCG) structures can be designed with a host of useful properties,
including reflectors~\cite{huang_surface-emitting_2007,bruckner_realization_2010}, focusing elements~\cite{lu_planar_2010,fattal_flat_2010}, filters, polarizers,  and resonators~\cite{chang-hasnain_high-contrast_2011}.  In earlier work, we 
incorporated a free-standing HCG in silicon nitride into a Fabry-Perot cavity and obtained a finesse  $F > 2800$, and found mechanical quality factors in the device as high as $Q\approx780\,000$~\cite{Kemiktarak2012}. 

Here we employ an HCG fabricated in a mechanically compliant silicon nitride membrane to realize a ``mirror in the middle'' cavity system with a membrane of high reflectivity,
as depicted in~Fig.~\ref{fig:matrix}a. 
We show how the mode structure rapidly changes near the points where the left sub-cavity and right sub-cavity simultaneously come into resonance (Fig.~\ref{fig:matrix}b), and suggest that this is best understood via a perturbation theory starting from unit reflectivity, in contrast to the usual dispersive regime for membrane-in-the-middle work.  In addition, the spectral signatures of the system allow  more detailed study of the losses than is possible in a simple cavity, and we quantify the reflection, transmission, absorption and scattering losses in the context of a simple model.

\section{HCG design and fabrication}
The HCG design was driven by the desire for a structure that would be relatively insensitive to fabrication parameters
while offering high reflectivity.  Although analytic formulations for HCG properties are available for idealized geometries~\cite{karagodsky_theoretical_2010},
we have relied on rigorous coupled wave analysis~\cite{moharam_rigorous_1981} (RCWA), which allows more general structures to be treated~\cite{Kemiktarak2012}. 
For low-stress silicon nitride, with an index of refraction of $n\approx2.2$, a thickness of
$t=470$~nm is found to allow high reflectivity at our target wavelength of 1560~nm with grating periods in the range
of $1.44\,\mu$m to $1.54\,\mu$m and finger widths in the range of  $0.52\,\mu$m to $0.62\,\mu$m.  In earlier work~\cite{stambaugh2014},
we found the imaginary part of the index of refraction of our silicon nitride membranes to be in the range $1.66\times10^{-5}<n_I<1.92\times10^{-5}$.
Taking $n_I=1.8\times10^{-5}$, we calculate the absorption of the HCG structure to be in the range of $1.3\times10^{-4}$ to $1.5\times10^{-4}$ when it is
illuminated from one side, depending on the exact geometry.  The unpatterned membrane, on the other hand, is found to have an absorption of $6.3\times10^{-5}$; the difference is the consequence of electric field enhancement inside the HCG structure.

The HCG fabrication procedure has been described in our previous work~\cite{Kemiktarak2012, Kemiktarak2012a}.
The size of the membrane used here is $250\; \mu\mathrm{m} \times 250 \;\mu\mathrm{m}$; as shown in Fig.~\ref{fig:matrix}c, 
the HCG is located near the center of the membrane, and is circular with a diameter of $80\;\mu\mathrm{m}$. 

\section{Model for ``HCG in the middle'' system}
\begin{figure}
\centering
\includegraphics{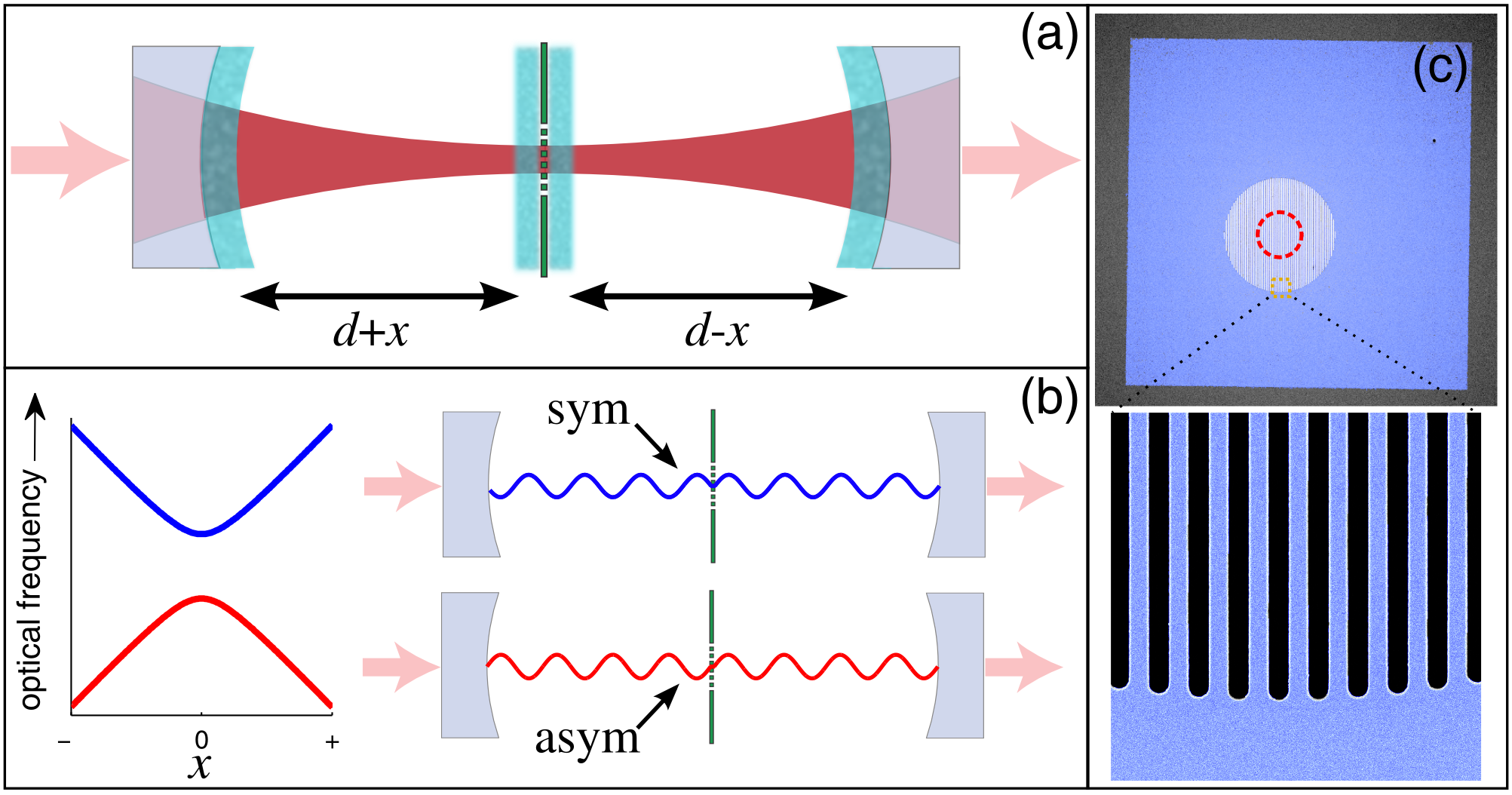}
\caption{(a) Model of experiment. The highly-reflective membrane is located near the center of an optical cavity.  It is represented
as a zero-thickness slab of polarizable material sandwiched between two ``scattering'' elements (light blue rectangles), each of which attenuates
the field in an optical traveling wave by $e^{-S_m/4}$.   The dielectric cavity mirrors are modeled as lossless ($R+T=1$) reflecting elements,
next to which similar ``scattering'' elements, characterized by $S_{diel}$, are placed.  The cavity is nearly concentric, resulting in a small beam waist. 
(b) Normal mode splitting. In the vicinity of membrane positions $x$ such that left and right sub-cavities are simultaneously resonant, the cavity
modes split into a doublet whose elements have opposite parity about the membrane.  (c) SEM images of a silicon nitride membrane (250~$\mu$m x 250~$\mu$m) with pattered HCG. The HCG has a diameter of 80~$\mu$m, and the small red circle represents the size (waist $\omega_0=17\,\mu$m) of the confined cavity mode. }
\label{fig:matrix}
\end{figure}

We model the ``HCG in the middle'' system by means of the transfer matrix formalism, which yields the
steady-state solution for the fields within and outside the cavity~\cite{Fowles}.  Each element
in the cavity is represented as a two-port device, in which the complex amplitudes of the outgoing and
ingoing electric fields on the right are related to the ingoing and outgoing fields on the left
by means of a matrix $M$, as follows:
\[
\left(
\begin{array}{c}
E_{out} \\
E_{in}
\end{array}
\right)_R
=M
\left(
\begin{array}{c}
E_{in} \\
E_{out}
\end{array}
\right)_L\,.
\]
The cavity length is denoted $2d$, and the membrane is positioned a distance $d+x$ from the input coupler (left mirror) of the cavity,
as shown in Fig.~\ref{fig:matrix}a.
The transfer matrix  $M^{cav}$ for the entire cavity can be found by simple matrix multiplication of the transfer matrices
of the individual elements, and the cavity transmission and reflection coefficients are given by
$t_{cav} =  1/M^{cav}_{22}$ and  $r_{cav} =  M^{cav}_{21}/M^{cav}_{22}$. 

The starting point for our description of the optical properties of the membrane is
that taken by Spencer and Lamb~\cite{Spencer_Lamb_PRA_1972} and others~\cite{fader_theory_1985,chow_composite-resonator_1986} 
in early studies of two coupled lasers, as well as more recent theoretical studies and proposals in 
optomechanics~\cite{bhattacharya_optomechanical_2008,xuereb_collectively_2013,Xuereb2012}.  The membrane is taken to be 
a zero-thickness slab of dielectric material with complex polarizability $\zeta = \zeta_R + i \zeta_I$, such that the
reflection and transmission coefficients are $r'_m=i\zeta/(1-i\zeta)$ and 
$t'_m=1/(1-i\zeta)$, respectively.  The corresponding transfer matrix is~\cite{xuereb_collectively_2013} 
\begin{equation}
M_{mem} = \left (
\begin{array}{cc}
1+ i \zeta & i \zeta \\ 
-i \zeta & 1-i \zeta 
\end{array}\right )\,.
\label{eqn: Mmem}
\end{equation}
Material absorption is described by $\zeta_I>0$; it is readily shown
that $|r'_m|^2+|t'_m|^2=1$ if and only if $\zeta_I=0$. While this model is clearly an idealization of the HCG, neglecting
its thickness and resonant properties, it does capture the essence of the device for these purposes.  

Nevertheless, a somewhat more general model is required to 
account for loss mechanisms other than absorption, such as scattering from surface roughness.  Indeed, 
if we consider light incident from only one side of the membrane described by~(\ref{eqn: Mmem}),
the fraction of the incident power lost to absorption is simply $A_m=1-|r'_m|^2-|t'_m|^2$.
If, however, light is incident from both sides,
of the same amplitude but differing in phase by~$\theta$, the fraction
of the incident power lost to absorption is found to be
\begin{equation*}
\frac{P_{in}-P_{out}}{P_{in}}=\frac{2\zeta_I(1+\cos\theta)}{\zeta_R^2+(1+\zeta_I)^2}\,.
\end{equation*}
The power absorbed vanishes even if $\zeta_I\ne0$ for $\theta=\pi$, meaning that the 
membrane is located at a node of the electromagnetic field.  In order
to model losses from other mechanisms, we sandwich the membrane between
two ``scattering'' elements (Fig.~\ref{fig:matrix}a), each of whose transfer matrix is taken to be
\begin{equation}
M_{scat} = \left (
\begin{array}{cc}
e^{-S_m/4} & 0 \\ 
0 & e^{S_m/4}
\end{array}\right ),
\end{equation}
where $S_m<<1$. The effect of this matrix is to attenuate the transmitted field by a factor $e^{-S_m/4}$ in a single pass, and thus the power by a factor $e^{-S_m/2}$.  For single-sided illumination, then, the fractions of the incident power reflected and transmitted are $R_m=|r'_m|^2e^{-S_m}$ and
$T_m=|t'_m|^2e^{-S_m}$, respectively, and one finds $R_m+T_m+A_m+S_m=1$.

For simplicity, we use the latter formalism to describe losses in the cavity mirrors as well.  
We model each dielectric cavity mirror as a lossless reflector, with real reflection and transmission coefficients $r_i$ and $t_i$ ($i=1,2$) satisfying $r_i^2+t_i^2=1$,
next to which a similar scattering element parameterized by $S_{diel}$ is located (Fig.\ref{fig:matrix}(a)).  

\section{Experiment}
\subsection{Setup}
The optical cavity used here is comprised of two high-reflectivity dielectric mirrors with nominal radii of curvature  
${\cal R}_{diel}=25\;\mathrm{mm}$.  The cavity length is set such that the optical cavity is nearly concentric, with a cavity length approximately $28 \pm 6\;\mathrm{\mu m}$ below the stability boundary of $2{\cal R}_{diel}$~\cite{kogelnik_laser_1966}.  
The geometry of the confined mode thus has a spot size (radius) on the mirrors of  $\omega_1=730\pm40\;\mathrm{\mu m}$ and a waist (radius) of $\omega_0=17\pm1\;\mathrm{\mu m}$; the numeric value following the $\pm$ sign is the combined standard uncertainty given with a confidence level of approximately $68\%$. This choice of waist is motivated by a compromise between minimizing beam spillage off of the patterned grating, and having a waist whose
wavevectors have a narrow distribution of transverse momenta, as noted previously~\cite{Kemiktarak2012a}.

The membrane is placed on a X-Y-Z stage with tip/tilt control so that it can be positioned at the center of the optical cavity. 
It is adjusted to be normal to the cavity mode and located at the mode waist by monitoring the transmission, and minimizing signatures
of coupling from the $\mathrm{TEM}_{00}$ mode to higher-order transverse modes.  The longitudinal location is confirmed by blocking the portion of the cavity following
the membrane so as to establish a simple Fabry-Perot cavity with the input coupler and the HCG, and measuring its free spectral range (FSR).  The FSR of this 
sub-cavity is 
a factor of two larger than that of the empty cavity (membrane removed); accounting for measurement uncertainty, we infer that the membrane is within $2.3\;\mathrm{\mu m}$ of the center of the cavity.


\subsection{Empty cavity}
\begin{figure}
\centering
\includegraphics{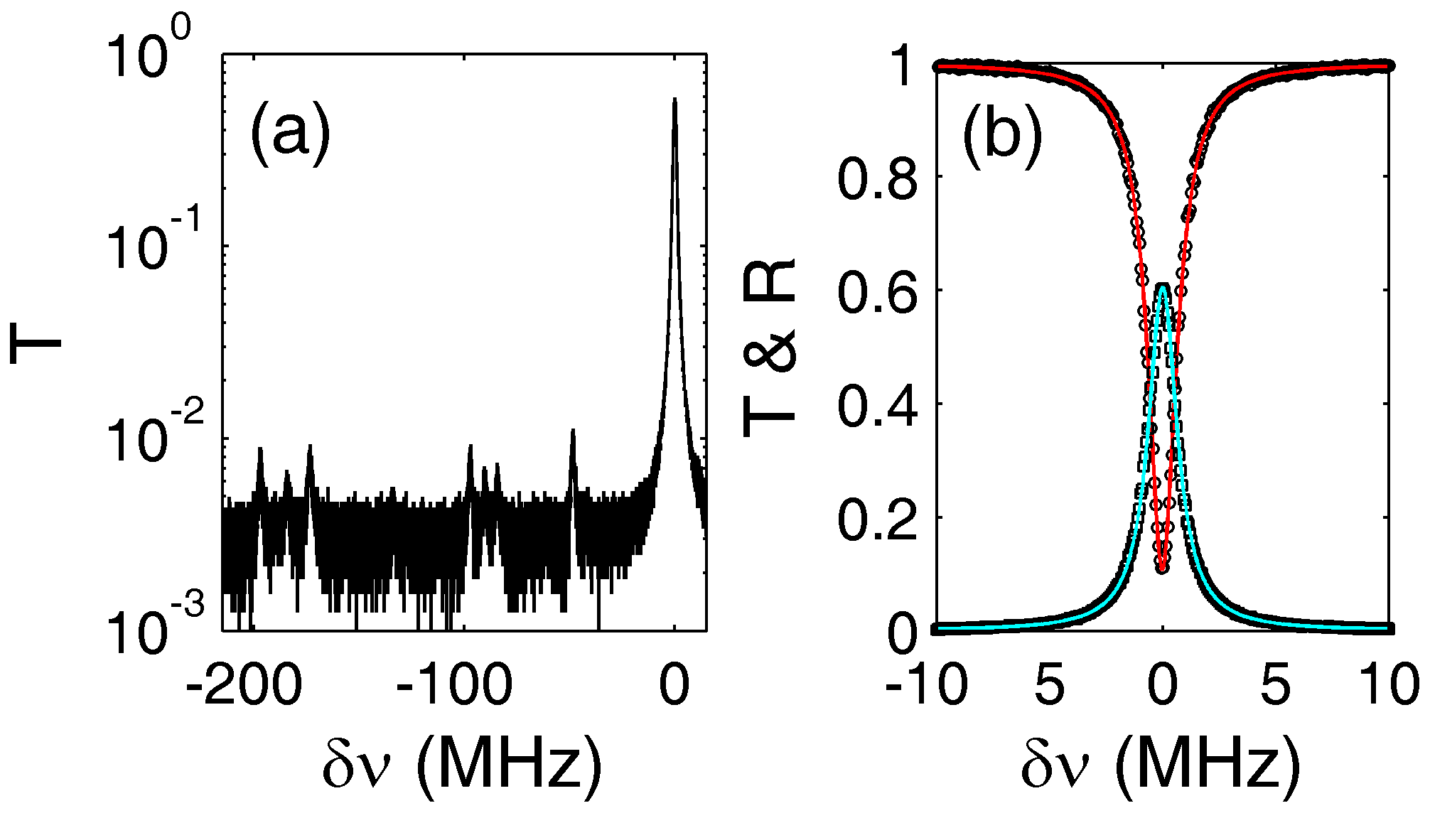}
\caption{(a) Transmission spectrum of the empty cavity, logarithmic scale. The large peak corresponds to a $\mathrm{TEM}_{00}$ mode, and the small peaks are higher-order transverse modes. Their small size relative to the main peak is indicative of good mode matching. (b) Transmission (cyan) and reflection (red) for the empty cavity,
with Lorentzian fits. From these data the characteristics of the cavity mirrors can be determined. }
\label{fig:figure1}
\end{figure}

We begin by studying the cavity in the absence of an HCG in order to establish the characteristics of the cavity mirrors.
In Fig.~\ref{fig:figure1}(a) the measured transmission of the empty cavity is plotted on a logarithmic scale. The large peak is the transmission of a fundamental $\mathrm{TEM}_{00}$
mode, and sets the origin of the detuning.  The small peaks, with amplitudes two orders of magnitude below that of the $\mathrm{TEM}_{00}$ mode, are due to
coupling to higher-order transverse cavity modes.  Their small size indicates that the injected beam is well mode-matched
to the cavity.  Moreover, their frequencies are related to the overall
cavity length and geometry of the confined modes~\cite{kogelnik_laser_1966}; in particular, one infers that the waist size of the fundamental mode is
$\omega_0=17\,\mu\mathrm{m}$, as noted previously.

In Fig.~\ref{fig:figure1}(b), the calibrated cavity transmission and reflection of a $\mathrm{TEM}_{00}$ resonance are plotted on a linear scale.  
The frequency scale was established by means of an auxiliary experiment using sidebands on the laser provided by an electro-optic modulator.
The predicted transmission and reflection spectra,taking $\zeta\rightarrow 0$ in our model, are
\begin{eqnarray}
T_{cav}&=&\frac{T_1T_2}{(S_{diel}+T_{avg})^2+\left(\frac{2\pi\delta\nu}{\Delta\nu}\right)^2} \label{eqn: Temptycav} \\
R_{cav}&=&1-\frac{T_1(T_2+2S_{diel})}{(S_{diel}+T_{avg})^2+\left(\frac{2\pi\delta\nu}{\Delta\nu}\right)^2}. \label{eqn: Remptycav} 
\end{eqnarray}
Here $T_1=t_1^2$ is the (power) transmission of the input coupler, $T_2=t_2^2$ is the transmission of the output coupler, $T_{avg}=(T_1+T_2)/2$, 
$\Delta\nu\equiv c/(4d)$ is the free spectral range of the cavity (length $2d$), and $\delta\nu=\nu-\nu_0$
is the detuning of the incident light with frequency $\nu$ from the cavity resonance frequency $\nu_0$.
The FWHM of the transmission peak is
\begin{equation}
\frac{\kappa}{2\pi}=\frac{S_{diel}+T_{avg}}{\pi}\Delta\nu. \label{eqn: FWHMemptycav}
\end{equation}
By fitting the data in~Fig.~\ref{fig:figure1}(b) to~(\ref{eqn: Temptycav}) and~(\ref{eqn: Remptycav}), we determine $S_{diel}=(3.6\pm0.1)\times10^{-4}$, $T_L=(1.07\pm0.04)\times10^{-3}$ and $T_R=(1.45\pm0.04)\times10^{-3}$. Having established the characteristics of the empty cavity, we are now in a position to study the cavity with an HCG in the middle.

\subsection{HCG in cavity}
\begin{figure}[]
\centering
\includegraphics{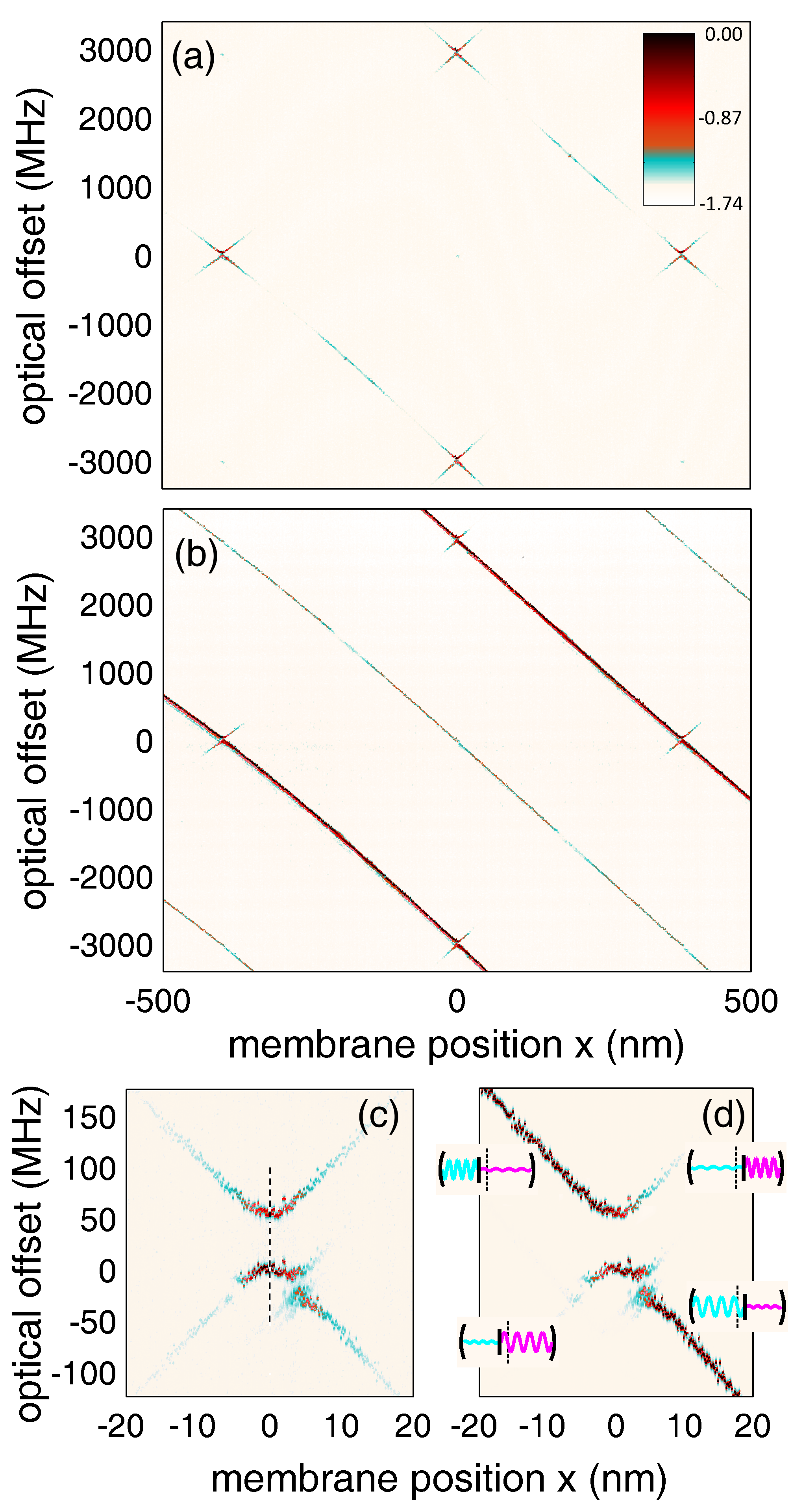}
\caption{Transmission and reflection spectra for membrane points $x$ near the center of the cavity. (a) The transmission is large only at points $x$
where both left and right sub-cavities can be simultaneously resonant.  The color scheme is logarithmic, with the colorbar representing $\log_{10}(T)$. 
(b) Here the color scheme represents $\log_{10}(1-R)$.  The reflection exhibits a pronounced dip for all frequencies such that the left 
(input) sub-cavity is resonant.  The weaker reflection dip arises from $\mathrm{TEM}_{10}$ modes
that have not been suppressed.  (c) Fine scan of transmission near the normal mode splitting. The crossing seen at $x\approx4$~nm results from the $\mathrm{TEM}_{02}$ 
symmetric mode coupling to the antisymmetric $\mathrm{TEM}_{00}$ mode. The dashed line at $x=0$ indicates the data slice plotted in Fig.~\ref{fig:figure4}a. 
(d) Zoom in on the reflection data; the insets represent the field distribution. The illustrations near the upper (lower) branch are associated with the symmetric  
(antisymmetric) modes.}
\label{fig:figure2}
\end{figure}

When a highly reflective HCG membrane is placed in the center of the cavity, the system behaves as two sub-cavities that are optically coupled
through the membrane transmission and mechanically coupled through the position of the membrane.  The transmission and reflection spectra are then functions
of the axial membrane position, and are shown in Fig.~\ref{fig:figure2}.
The transmission, Fig.~\ref{fig:figure2}a, is largely maximized when each sub-cavity is simultaneously resonant,
which is possible when $x\approx N\lambda/4$. The reflection, on the other hand, exhibits a deep dip whenever the left-hand
sub-cavity is resonant, as shown in Fig.~\ref{fig:figure2}b. Qualitatively, this can be understood as the left-hand sub-cavity
being approximately impedance matched~\cite{Siegman1986}.  At the points where the transmission is maximal, there are in fact
pronounced avoided crossings in the spectra, as shown in Figures~\ref{fig:figure2}c and~\ref{fig:figure2}d.
(The distortion of the lower branch of the resonance curves at $x\approx 5$~nm is due to the presence of higher-order transverse modes,
and will be pursued in a subsequent publication).  At the avoided crossing at $x=0$, the lower resonance corresponds to an
optical cavity mode with odd parity, and the upper resonance corresponds to an optical mode with even parity (Fig.~\ref{fig:matrix}b).  
For $x\ne 0$, the mode amplitudes in the two sub-cavities are different, and the modes are no longer purely symmetric or antisymmetric. 
For $-\lambda/8<x<0$, the higher-frequency mode is localized primarily in the left sub-cavity,
while for  $0<x<\lambda/8$, it is the lower-frequency mode that is localized to the left.  
This is illustrated in the close-up of the reflection spectrum shown in 
Fig.~\ref{fig:figure2}d. Analytic expressions for the modes of a lossless cavity, with perfectly reflecting cavity mirrors and $\zeta_I\rightarrow 0$, have been given
previously~\cite{fader_theory_1985,chow_composite-resonator_1986}.

In our model of a zero-thickness membrane, the field amplitude of the antisymmetric mode vanishes at the membrane location, and the wavelength 
of the mode is the same as that of the empty cavity at the same frequency.  The symmetric mode, however, is 
nonvanishing, with a discontinuous spatial derivative at the membrane position~\cite{fader_theory_1985}.  As
$R_m\equiv|r_m|^2\rightarrow1$, the field amplitude at the membrane diminishes and approaches a node, but as long as $R_m<1$ 
the amplitude will be nonzero and result in absorption.  Similarly, the phase accumulation
from $-d$ to $d$ remains larger than that for the antisymmetric mode, with the consequence that the frequency is higher.
We denote the frequency difference between the symmetric and antisymmetric modes at $x=0$ as $\delta\nu_{SA}$.

\begin{figure}
\centering
\includegraphics{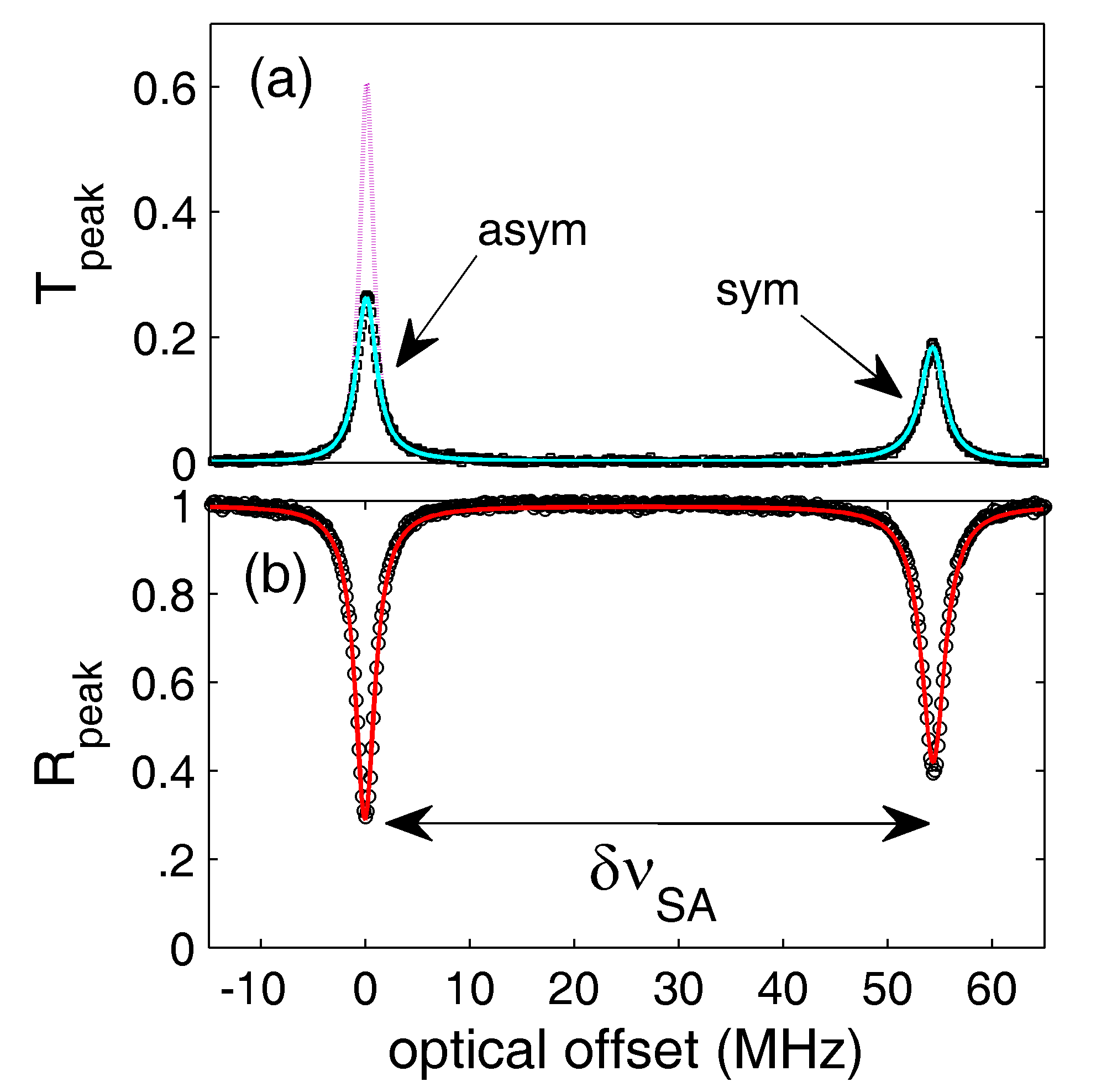}
\caption{(a) Transmission and (b) reflection at $x=0$ (dashed line in Fig.~\ref{fig:figure2}c). The two peaks correspond to the antisymmetric and symmetric modes. The taller peak (purple) in (a) is the empty cavity transmission. The overall reduction in height of both modes with respect to the empty cavity results from scattering losses, while the relatively smaller, broader peak of the symmetric mode results from absorption. The frequency separation $\delta\nu_{SA}$ is used to determine $\left|t_m \right|$ or $\zeta_R$. The green curve superposed on the transmission data is a fit to our model, in which the only adjusted parameters are the loss terms $S_m$ and $\zeta_I$.
The red curve superposed on the reflection data is not a fit, but rather the prediction of our model using the parameters $S_m$ and $\zeta_I$ determined from the fit to the transmission data.}
\label{fig:figure4}
\end{figure}

The transmission and reflection spectra, accounting for cavity and membrane losses, are given by our transfer matrix treatment.   
Here we focus on the case $x=0$.  
The transmission spectrum at $x=0$ is shown in Fig.~\ref{fig:figure4}a, where the origin of the detuning is taken to be
the resonant frequency of the antisymmetric mode.  The mode splitting is seen to be $\delta\nu_{SA}=54.16\pm0.22$~MHz, and the resonance associated
with the antisymmetric mode is somewhat stronger and narrower than that associated with the symmetric mode.  From the mode splitting we can
infer the membrane transmission.  Within our model, the mode splitting is independent of scattering losses and is given by
\begin{eqnarray}
\delta \nu_{SA}&=& \frac{ \Delta \nu}{\pi} \tan^{-1}\left(\frac{2\zeta_R}{\zeta_R^2-1}\left(1-\frac{\zeta_I^2}{\zeta_R^2-1}\right)\right) \label{eq:exactsplit} \\
&\approx& \frac{2}{\pi} (\sin^{-1}|t_m|) \Delta \nu, \label{eq:split}
\end{eqnarray} 
where the last line results from making the approximation of a membrane without absorption loss, $\zeta_I\rightarrow 0$.
This result agrees with that found in earlier work~\cite{fader_theory_1985,chow_composite-resonator_1986,bhattacharya_optomechanical_2008}. From~(\ref{eq:split})
we infer $T_m=(8.07\pm0.06)\times10^{-4}$, with a corresponding polarizability of $\zeta=35.18\pm0.14$. 
The correction to $\zeta_R$ induced by taking $\zeta_I$ (as determined from the subsequent lineshape analysis) in~(\ref{eq:exactsplit}) is negligible.

We next consider the lineshapes of the transmission resonances shown in Fig.~\ref{fig:figure4}a.  
In the interests of transparency, we give expressions for a symmetric cavity ($T_1=T_2\equiv T$), although we make
use of the more general situation ($T_1\ne T_2$, with $T_1$ and $T_2$ determined from our study of the empty cavity) in our subsequent analysis.
The transmission spectrum of the antisymmetric mode
is found to be Lorentzian with FWHM and peak height given by
\begin{eqnarray}\label{eq:gamma-node}
\frac{\kappa^{asym}}{2\pi} &=& \frac{S+T}{\pi}\Delta\nu \label{eq:gamma-node} \\
T^{asym}_{peak}&=&\frac{T^2}{\left(\left(1+\frac{(T+S)^2}{16}\zeta_R^2\right)+\frac{T+S}{2}\zeta_I\right)(T+S)^2}, \label{eq:t-node}
\end{eqnarray}
where $S=S_{diel}+S_m$. The expression for the linewidth is the same as~(\ref{eqn: FWHMemptycav}) for the empty cavity,
with the scattering losses now taken to be the sum of those for the membrane and the dielectric mirrors.

The FWHM and peak height of the symmetric mode are given by
\begin{eqnarray}
\frac{\kappa^{sym}}{2\pi} &=&\frac{1}{\pi}\sqrt{\frac{(S+T)(8 \zeta_i+(S+T)(1+\zeta_R^2))}{1+\zeta_R^2}}\Delta\nu\label{eq:gamma-anode} \\
T^{sym}_{peak}&=&\frac{T^2}{(\frac{T+S}{2})^2+\left(\pi\frac{\delta\nu_{SA}}{\Delta\nu}\right)^2} \nonumber \\
&&\times\frac{1}{2\zeta_I (S+T)+(1+\zeta_R^2)\left(\frac{T+S}{2}\right)^2}, \label{eq:t-anode}
\end{eqnarray}
where it has been assumed that $\zeta_R>>1$ (which is the case of interest).
If  $\zeta_I\ne0$ the peak is not exactly Lorentzian, and it shows additional broadening due to absorption.

Taking the mirror properties $T_1$, $T_2$ and $S_{diel}$ from the measurements of the empty cavity, and $\delta\nu_{SA}$ and $\zeta_R$ from
the measured frequency splitting between the symmetric and antisymmetric modes~(\ref{eq:split}), the only
remaining parameters to be determined are the membrane scattering $S_m$ and the imaginary part of the 
polarizability~$\zeta_I$.  Fitting the transmission data, as shown in Fig.~\ref{fig:figure4}a, to generalizations of~(\ref{eq:gamma-node})-(\ref{eq:t-anode})
for $T_1\ne T_2$ yields $\zeta_I=0.145\pm0.008$ and $S_m=(8 \pm 1)\times10^{-4}$. These parameters may then be substituted into expressions for the 
cavity reflection based on the matrix model, and the results are overlaid on the reflection data in Fig.~\ref{fig:figure4}b;
the excellent agreement is a testament that our model accurately captures the underlying physics.

We now summarize the optical properties of the HCG as determined from our measurements.  Within our model, the reflection and transmission
of the slab parametrized by polarizability $\zeta$ are $|r'_m|^2=|i\zeta/(1-i\zeta)|^2=0.99896\pm2\times10^{-5}$ and $|t'_m|^2=|1/(1-i\zeta)|^2=(8.07\pm0.06)\times10^{-4}$.
The corresponding absorption, appropriate for light incident from one side of the HCG, is $A_m=1-|r'_m|^2-|t'_m|^2=(2.34\pm0.28)\times10^{-4}$.  The transmission
is thus approximately a factor of 3.5 higher than the absorption.  Scattering
reduces the transmission and reflection by a factor $e^{-S_m}$, yielding $R_m=|r'_m|^2e^{-S_m}=0.9982$ and $T_m=|t'_m|^2e^{-S_m}=8.06\times10^{-4}$.

\section{Conclusion}
By using a membrane patterned as a subwavelength diffraction grating, we have implemented a ``membrane in the middle'' system with a membrane of very high reflectivity.  The resonance spectrum is qualitatively different from those demonstrated to date using low-reflectivity membranes; rather than a sinusoidal modulation of frequency with membrane displacement, the spectrum is that of two cavities coupled by photon tunneling through a shared highly-reflecting mirror.  The weak optical coupling manifests itself as an avoided crossing at the point where the left- and right-hand cavities are simultaneously resonant, and from the size of the avoided crossing we are able to precisely determine the membrane transmission.

Analysis of the avoided crossing allows additional insight into the loss mechanisms present in the subwavelength grating.  The upper branch of the avoided crossing is a spatially symmetric mode, having nonzero overlap with the membrane, whereas the lower branch is antisymmetric, having (in the limit of a membrane of vanishing thickness) no overlap.  Consequently, the symmetric mode suffers greater loss from absorption, manifest in the form of a weaker resonance peak with a larger linewidth.  
Losses unrelated to material absorption are the same for both modes, so that by comparing the two resonance peaks with that of the empty cavity, we are able to distinguish the effects of absorption and scattering.  


The values found for the HCG properties agree reasonably well with those expected from RCWA calculations, although the measured absorption
$A_m=2.34\times10^{-4}$ is somewhat higher than the value $1.3\times10^{-4}<A<1.5\times10^{-4}$ predicted by RCWA.  Further reduction of absorption losses
is possible by the use of stoichiometric Si$_3$N$_4$~\cite{Wilson2009} rather than the low-stress SiN that was used in this work.  
Scattering losses can in principle be reduced by further improving the fabrication process.  The present study has concentrated exclusively on the
optical properties of our ``HCG in the middle'' system.  The HCG is, however, mechanically compliant, and future work will explore the optomechanical
opportunities in this system.

\section{Acknowledgments}
Research performed in part at the NIST Center for Nanoscale Science and Technology.

\bibliography{article}

\begin{thebibliography}{42}%
\makeatletter
\providecommand \@ifxundefined [1]{%
 \@ifx{#1\undefined}
}%
\providecommand \@ifnum [1]{%
 \ifnum #1\expandafter \@firstoftwo
 \else \expandafter \@secondoftwo
 \fi
}%
\providecommand \@ifx [1]{%
 \ifx #1\expandafter \@firstoftwo
 \else \expandafter \@secondoftwo
 \fi
}%
\providecommand \natexlab [1]{#1}%
\providecommand \enquote  [1]{``#1''}%
\providecommand \bibnamefont  [1]{#1}%
\providecommand \bibfnamefont [1]{#1}%
\providecommand \citenamefont [1]{#1}%
\providecommand \href@noop [0]{\@secondoftwo}%
\providecommand \href [0]{\begingroup \@sanitize@url \@href}%
\providecommand \@href[1]{\@@startlink{#1}\@@href}%
\providecommand \@@href[1]{\endgroup#1\@@endlink}%
\providecommand \@sanitize@url [0]{\catcode `\\12\catcode `\$12\catcode
  `\&12\catcode `\#12\catcode `\^12\catcode `\_12\catcode `\%12\relax}%
\providecommand \@@startlink[1]{}%
\providecommand \@@endlink[0]{}%
\providecommand \url  [0]{\begingroup\@sanitize@url \@url }%
\providecommand \@url [1]{\endgroup\@href {#1}{\urlprefix }}%
\providecommand \urlprefix  [0]{URL }%
\providecommand \Eprint [0]{\href }%
\providecommand \doibase [0]{http://dx.doi.org/}%
\providecommand \selectlanguage [0]{\@gobble}%
\providecommand \bibinfo  [0]{\@secondoftwo}%
\providecommand \bibfield  [0]{\@secondoftwo}%
\providecommand \translation [1]{[#1]}%
\providecommand \BibitemOpen [0]{}%
\providecommand \bibitemStop [0]{}%
\providecommand \bibitemNoStop [0]{.\EOS\space}%
\providecommand \EOS [0]{\spacefactor3000\relax}%
\providecommand \BibitemShut  [1]{\csname bibitem#1\endcsname}%
\let\auto@bib@innerbib\@empty
\bibitem [{\citenamefont {Aspelmeyer}\ \emph {et~al.}(2013)\citenamefont
  {Aspelmeyer}, \citenamefont {Kippenberg},\ and\ \citenamefont
  {Marquardt}}]{Aspelmeyer2013}%
  \BibitemOpen
  \bibfield  {author} {\bibinfo {author} {\bibfnamefont {M.}~\bibnamefont
  {Aspelmeyer}}, \bibinfo {author} {\bibfnamefont {T.~J.}\ \bibnamefont
  {Kippenberg}}, \ and\ \bibinfo {author} {\bibfnamefont {F.}~\bibnamefont
  {Marquardt}},\ }\href@noop {} {\bibfield  {journal} {\bibinfo  {journal}
  {arxiv:1303.0733}\ } (\bibinfo {year} {2013})},\ \Eprint
  {http://arxiv.org/abs/1303.0733} {1303.0733} \BibitemShut {NoStop}%
\bibitem [{\citenamefont {O'Connell}\ \emph {et~al.}(2010)\citenamefont
  {O'Connell}, \citenamefont {Hofheinz}, \citenamefont {Ansmann}, \citenamefont
  {Bialczak}, \citenamefont {Lenander}, \citenamefont {Lucero}, \citenamefont
  {Neeley}, \citenamefont {Sank}, \citenamefont {Wang}, \citenamefont {Weides},
  \citenamefont {Wenner}, \citenamefont {Martinis},\ and\ \citenamefont
  {Cleland}}]{OConnell2010}%
  \BibitemOpen
  \bibfield  {author} {\bibinfo {author} {\bibfnamefont {A.~D.}\ \bibnamefont
  {O'Connell}}, \bibinfo {author} {\bibfnamefont {M.}~\bibnamefont {Hofheinz}},
  \bibinfo {author} {\bibfnamefont {M.}~\bibnamefont {Ansmann}}, \bibinfo
  {author} {\bibfnamefont {R.~C.}\ \bibnamefont {Bialczak}}, \bibinfo {author}
  {\bibfnamefont {M.}~\bibnamefont {Lenander}}, \bibinfo {author}
  {\bibfnamefont {E.}~\bibnamefont {Lucero}}, \bibinfo {author} {\bibfnamefont
  {M.}~\bibnamefont {Neeley}}, \bibinfo {author} {\bibfnamefont
  {D.}~\bibnamefont {Sank}}, \bibinfo {author} {\bibfnamefont {H.}~\bibnamefont
  {Wang}}, \bibinfo {author} {\bibfnamefont {M.}~\bibnamefont {Weides}},
  \bibinfo {author} {\bibfnamefont {J.}~\bibnamefont {Wenner}}, \bibinfo
  {author} {\bibfnamefont {J.~M.}\ \bibnamefont {Martinis}}, \ and\ \bibinfo
  {author} {\bibfnamefont {A.~N.}\ \bibnamefont {Cleland}},\ }\href
  {http://dx.doi.org/10.1038/nature08967} {\bibfield  {journal} {\bibinfo
  {journal} {Nature}\ }\textbf {\bibinfo {volume} {464}},\ \bibinfo {pages}
  {697} (\bibinfo {year} {2010})}\BibitemShut {NoStop}%
\bibitem [{\citenamefont {Brooks}\ \emph {et~al.}(2012)\citenamefont {Brooks},
  \citenamefont {Botter}, \citenamefont {Schreppler}, \citenamefont {Purdy},
  \citenamefont {Brahms},\ and\ \citenamefont {Stamper-Kurn}}]{Brooks2012}%
  \BibitemOpen
  \bibfield  {author} {\bibinfo {author} {\bibfnamefont {D.~W.~C.}\
  \bibnamefont {Brooks}}, \bibinfo {author} {\bibfnamefont {T.}~\bibnamefont
  {Botter}}, \bibinfo {author} {\bibfnamefont {S.}~\bibnamefont {Schreppler}},
  \bibinfo {author} {\bibfnamefont {T.~P.}\ \bibnamefont {Purdy}}, \bibinfo
  {author} {\bibfnamefont {N.}~\bibnamefont {Brahms}}, \ and\ \bibinfo {author}
  {\bibfnamefont {D.~M.}\ \bibnamefont {Stamper-Kurn}},\ }\href
  {http://dx.doi.org/10.1038/nature11325} {\bibfield  {journal} {\bibinfo
  {journal} {Nature}\ }\textbf {\bibinfo {volume} {488}},\ \bibinfo {pages}
  {476} (\bibinfo {year} {2012})}\BibitemShut {NoStop}%
\bibitem [{\citenamefont {Safavi-Naeini}\ \emph {et~al.}(2013)\citenamefont
  {Safavi-Naeini}, \citenamefont {Groblacher}, \citenamefont {Hill},
  \citenamefont {Chan}, \citenamefont {Aspelmeyer},\ and\ \citenamefont
  {Painter}}]{Safavi-Naeini2013}%
  \BibitemOpen
  \bibfield  {author} {\bibinfo {author} {\bibfnamefont {A.~H.}\ \bibnamefont
  {Safavi-Naeini}}, \bibinfo {author} {\bibfnamefont {S.}~\bibnamefont
  {Groblacher}}, \bibinfo {author} {\bibfnamefont {J.~T.}\ \bibnamefont
  {Hill}}, \bibinfo {author} {\bibfnamefont {J.}~\bibnamefont {Chan}}, \bibinfo
  {author} {\bibfnamefont {M.}~\bibnamefont {Aspelmeyer}}, \ and\ \bibinfo
  {author} {\bibfnamefont {O.}~\bibnamefont {Painter}},\ }\href
  {http://dx.doi.org/10.1038/nature12307} {\bibfield  {journal} {\bibinfo
  {journal} {Nature}\ }\textbf {\bibinfo {volume} {500}},\ \bibinfo {pages}
  {185} (\bibinfo {year} {2013})}\BibitemShut {NoStop}%
\bibitem [{\citenamefont {Purdy}\ \emph
  {et~al.}(2013{\natexlab{a}})\citenamefont {Purdy}, \citenamefont {Yu},
  \citenamefont {Peterson}, \citenamefont {Kampel},\ and\ \citenamefont
  {Regal}}]{purdy_strong_2013}%
  \BibitemOpen
  \bibfield  {author} {\bibinfo {author} {\bibfnamefont {T.~P.}\ \bibnamefont
  {Purdy}}, \bibinfo {author} {\bibfnamefont {P.-L.}\ \bibnamefont {Yu}},
  \bibinfo {author} {\bibfnamefont {R.~W.}\ \bibnamefont {Peterson}}, \bibinfo
  {author} {\bibfnamefont {N.~S.}\ \bibnamefont {Kampel}}, \ and\ \bibinfo
  {author} {\bibfnamefont {C.~A.}\ \bibnamefont {Regal}},\ }\href {\doibase
  10.1103/PhysRevX.3.031012} {\bibfield  {journal} {\bibinfo  {journal}
  {Physical Review X}\ }\textbf {\bibinfo {volume} {3}} (\bibinfo {year}
  {2013}{\natexlab{a}}),\ 10.1103/PhysRevX.3.031012}\BibitemShut {NoStop}%
\bibitem [{\citenamefont {Bhattacharya}\ \emph {et~al.}(2008)\citenamefont
  {Bhattacharya}, \citenamefont {Uys},\ and\ \citenamefont
  {Meystre}}]{bhattacharya_optomechanical_2008}%
  \BibitemOpen
  \bibfield  {author} {\bibinfo {author} {\bibfnamefont {M.}~\bibnamefont
  {Bhattacharya}}, \bibinfo {author} {\bibfnamefont {H.}~\bibnamefont {Uys}}, \
  and\ \bibinfo {author} {\bibfnamefont {P.}~\bibnamefont {Meystre}},\
  }\href@noop {} {\bibfield  {journal} {\bibinfo  {journal} {Phys. Rev. A}\
  }\textbf {\bibinfo {volume} {77}} (\bibinfo {year} {2008})}\BibitemShut
  {NoStop}%
\bibitem [{\citenamefont {Thompson}\ \emph {et~al.}(2008)\citenamefont
  {Thompson}, \citenamefont {Zwickl}, \citenamefont {Jayich}, \citenamefont
  {Marquardt}, \citenamefont {Girvin},\ and\ \citenamefont
  {Harris}}]{Thompson2008a}%
  \BibitemOpen
  \bibfield  {author} {\bibinfo {author} {\bibfnamefont {J.~D.}\ \bibnamefont
  {Thompson}}, \bibinfo {author} {\bibfnamefont {B.~M.}\ \bibnamefont
  {Zwickl}}, \bibinfo {author} {\bibfnamefont {A.~M.}\ \bibnamefont {Jayich}},
  \bibinfo {author} {\bibfnamefont {F.}~\bibnamefont {Marquardt}}, \bibinfo
  {author} {\bibfnamefont {S.~M.}\ \bibnamefont {Girvin}}, \ and\ \bibinfo
  {author} {\bibfnamefont {J.~G.~E.}\ \bibnamefont {Harris}},\ }\href {\doibase
  10.1038/nature06715} {\bibfield  {journal} {\bibinfo  {journal} {Nature}\
  }\textbf {\bibinfo {volume} {452}},\ \bibinfo {pages} {72} (\bibinfo {year}
  {2008})}\BibitemShut {NoStop}%
\bibitem [{\citenamefont {Purdy}\ \emph
  {et~al.}(2013{\natexlab{b}})\citenamefont {Purdy}, \citenamefont {Peterson},\
  and\ \citenamefont {Regal}}]{purdy_observation_2013}%
  \BibitemOpen
  \bibfield  {author} {\bibinfo {author} {\bibfnamefont {T.~P.}\ \bibnamefont
  {Purdy}}, \bibinfo {author} {\bibfnamefont {R.~W.}\ \bibnamefont {Peterson}},
  \ and\ \bibinfo {author} {\bibfnamefont {C.~A.}\ \bibnamefont {Regal}},\
  }\href {\doibase 10.1126/science.1231282} {\bibfield  {journal} {\bibinfo
  {journal} {Science}\ }\textbf {\bibinfo {volume} {339}},\ \bibinfo {pages}
  {801} (\bibinfo {year} {2013}{\natexlab{b}})}\BibitemShut {NoStop}%
\bibitem [{\citenamefont {Karuza}\ \emph {et~al.}(2013)\citenamefont {Karuza},
  \citenamefont {Biancofiore}, \citenamefont {Bawaj}, \citenamefont
  {Molinelli}, \citenamefont {Galassi}, \citenamefont {Natali}, \citenamefont
  {Tombesi}, \citenamefont {Di~Giuseppe},\ and\ \citenamefont
  {Vitali}}]{karuza_optomechanically_2013}%
  \BibitemOpen
  \bibfield  {author} {\bibinfo {author} {\bibfnamefont {M.}~\bibnamefont
  {Karuza}}, \bibinfo {author} {\bibfnamefont {C.}~\bibnamefont {Biancofiore}},
  \bibinfo {author} {\bibfnamefont {M.}~\bibnamefont {Bawaj}}, \bibinfo
  {author} {\bibfnamefont {C.}~\bibnamefont {Molinelli}}, \bibinfo {author}
  {\bibfnamefont {M.}~\bibnamefont {Galassi}}, \bibinfo {author} {\bibfnamefont
  {R.}~\bibnamefont {Natali}}, \bibinfo {author} {\bibfnamefont
  {P.}~\bibnamefont {Tombesi}}, \bibinfo {author} {\bibfnamefont
  {G.}~\bibnamefont {Di~Giuseppe}}, \ and\ \bibinfo {author} {\bibfnamefont
  {D.}~\bibnamefont {Vitali}},\ }\href {\doibase 10.1103/PhysRevA.88.013804}
  {\bibfield  {journal} {\bibinfo  {journal} {Phys. Rev. A}\ }\textbf {\bibinfo
  {volume} {88}} (\bibinfo {year} {2013}),\
  10.1103/PhysRevA.88.013804}\BibitemShut {NoStop}%
\bibitem [{\citenamefont {Shkarin}\ \emph {et~al.}(2014)\citenamefont
  {Shkarin}, \citenamefont {Flowers-Jacobs}, \citenamefont {Hoch},
  \citenamefont {Kashkanova}, \citenamefont {Deutsch}, \citenamefont
  {Reichel},\ and\ \citenamefont {Harris}}]{shkarin_optically_2014}%
  \BibitemOpen
  \bibfield  {author} {\bibinfo {author} {\bibfnamefont {A.}~\bibnamefont
  {Shkarin}}, \bibinfo {author} {\bibfnamefont {N.}~\bibnamefont
  {Flowers-Jacobs}}, \bibinfo {author} {\bibfnamefont {S.}~\bibnamefont
  {Hoch}}, \bibinfo {author} {\bibfnamefont {A.}~\bibnamefont {Kashkanova}},
  \bibinfo {author} {\bibfnamefont {C.}~\bibnamefont {Deutsch}}, \bibinfo
  {author} {\bibfnamefont {J.}~\bibnamefont {Reichel}}, \ and\ \bibinfo
  {author} {\bibfnamefont {J.}~\bibnamefont {Harris}},\ }\href {\doibase
  10.1103/PhysRevLett.112.013602} {\bibfield  {journal} {\bibinfo  {journal}
  {Phys. Rev. Lett.}\ }\textbf {\bibinfo {volume} {112}} (\bibinfo {year}
  {2014}),\ 10.1103/PhysRevLett.112.013602}\BibitemShut {NoStop}%
\bibitem [{\citenamefont {Xuereb}\ \emph {et~al.}(2013)\citenamefont {Xuereb},
  \citenamefont {Genes},\ and\ \citenamefont
  {Dantan}}]{xuereb_collectively_2013}%
  \BibitemOpen
  \bibfield  {author} {\bibinfo {author} {\bibfnamefont {A.}~\bibnamefont
  {Xuereb}}, \bibinfo {author} {\bibfnamefont {C.}~\bibnamefont {Genes}}, \
  and\ \bibinfo {author} {\bibfnamefont {A.}~\bibnamefont {Dantan}},\ }\href
  {\doibase 10.1103/PhysRevA.88.053803} {\bibfield  {journal} {\bibinfo
  {journal} {Phys. Rev. A}\ }\textbf {\bibinfo {volume} {88}} (\bibinfo {year}
  {2013}),\ 10.1103/PhysRevA.88.053803}\BibitemShut {NoStop}%
\bibitem [{\citenamefont {Xuereb}\ \emph {et~al.}(2012)\citenamefont {Xuereb},
  \citenamefont {Genes},\ and\ \citenamefont {Dantan}}]{Xuereb2012}%
  \BibitemOpen
  \bibfield  {author} {\bibinfo {author} {\bibfnamefont {A.}~\bibnamefont
  {Xuereb}}, \bibinfo {author} {\bibfnamefont {C.}~\bibnamefont {Genes}}, \
  and\ \bibinfo {author} {\bibfnamefont {A.}~\bibnamefont {Dantan}},\ }\href
  {\doibase 10.1103/PhysRevLett.109.223601} {\bibfield  {journal} {\bibinfo
  {journal} {Phys. Rev. Lett.}\ }\textbf {\bibinfo {volume} {109}},\ \bibinfo
  {pages} {223601} (\bibinfo {year} {2012})}\BibitemShut {NoStop}%
\bibitem [{\citenamefont {Genes}\ \emph {et~al.}(2013)\citenamefont {Genes},
  \citenamefont {Xuereb}, \citenamefont {Pupillo},\ and\ \citenamefont
  {Dantan}}]{genes_enhanced_2013}%
  \BibitemOpen
  \bibfield  {author} {\bibinfo {author} {\bibfnamefont {C.}~\bibnamefont
  {Genes}}, \bibinfo {author} {\bibfnamefont {A.}~\bibnamefont {Xuereb}},
  \bibinfo {author} {\bibfnamefont {G.}~\bibnamefont {Pupillo}}, \ and\
  \bibinfo {author} {\bibfnamefont {A.}~\bibnamefont {Dantan}},\ }\href
  {\doibase 10.1103/PhysRevA.88.033855} {\bibfield  {journal} {\bibinfo
  {journal} {Phys. Rev. A}\ }\textbf {\bibinfo {volume} {88}} (\bibinfo {year}
  {2013}),\ 10.1103/PhysRevA.88.033855}\BibitemShut {NoStop}%
\bibitem [{\citenamefont {Xu}\ and\ \citenamefont {Taylor}(2013)}]{Xu2013a}%
  \BibitemOpen
  \bibfield  {author} {\bibinfo {author} {\bibfnamefont {X.}~\bibnamefont
  {Xu}}\ and\ \bibinfo {author} {\bibfnamefont {J.~M.}\ \bibnamefont
  {Taylor}},\ }\href@noop {} {\bibfield  {journal} {\bibinfo  {journal}
  {arXiv:1303.7469}\ } (\bibinfo {year} {2013})}\BibitemShut {NoStop}%
\bibitem [{\citenamefont {Xu}\ \emph {et~al.}(2014)\citenamefont {Xu},
  \citenamefont {Gullans},\ and\ \citenamefont {Taylor}}]{Xu2013b}%
  \BibitemOpen
  \bibfield  {author} {\bibinfo {author} {\bibfnamefont {X.}~\bibnamefont
  {Xu}}, \bibinfo {author} {\bibfnamefont {M.}~\bibnamefont {Gullans}}, \ and\
  \bibinfo {author} {\bibfnamefont {J.~M.}\ \bibnamefont {Taylor}},\
  }\href@noop {} {\bibfield  {journal} {\bibinfo  {journal} {arxiv:1404.3726}\
  } (\bibinfo {year} {2014})},\ \Eprint {http://arxiv.org/abs/1404.3726}
  {1404.3726} \BibitemShut {NoStop}%
\bibitem [{\citenamefont {Sankey}\ \emph {et~al.}(2010)\citenamefont {Sankey},
  \citenamefont {Yang}, \citenamefont {Zwickl}, \citenamefont {Jayich},\ and\
  \citenamefont {Harris}}]{sankey_strong_2010}%
  \BibitemOpen
  \bibfield  {author} {\bibinfo {author} {\bibfnamefont {J.}~\bibnamefont
  {Sankey}}, \bibinfo {author} {\bibfnamefont {C.}~\bibnamefont {Yang}},
  \bibinfo {author} {\bibfnamefont {B.}~\bibnamefont {Zwickl}}, \bibinfo
  {author} {\bibfnamefont {A.}~\bibnamefont {Jayich}}, \ and\ \bibinfo {author}
  {\bibfnamefont {J.}~\bibnamefont {Harris}},\ }\href@noop {} {\bibfield
  {journal} {\bibinfo  {journal} {Nat. Phys.}\ }\textbf {\bibinfo {volume}
  {6}},\ \bibinfo {pages} {707} (\bibinfo {year} {2010})}\BibitemShut {NoStop}%
\bibitem [{\citenamefont {Flowers-Jacobs}\ \emph {et~al.}(2012)\citenamefont
  {Flowers-Jacobs}, \citenamefont {Hoch}, \citenamefont {Sankey}, \citenamefont
  {Kashkanova}, \citenamefont {Jayich}, \citenamefont {Deutsch}, \citenamefont
  {Reichel},\ and\ \citenamefont
  {Harris}}]{flowers-jacobs_fiber-cavity-based_2012}%
  \BibitemOpen
  \bibfield  {author} {\bibinfo {author} {\bibfnamefont {N.~E.}\ \bibnamefont
  {Flowers-Jacobs}}, \bibinfo {author} {\bibfnamefont {S.~W.}\ \bibnamefont
  {Hoch}}, \bibinfo {author} {\bibfnamefont {J.~C.}\ \bibnamefont {Sankey}},
  \bibinfo {author} {\bibfnamefont {A.}~\bibnamefont {Kashkanova}}, \bibinfo
  {author} {\bibfnamefont {A.~M.}\ \bibnamefont {Jayich}}, \bibinfo {author}
  {\bibfnamefont {C.}~\bibnamefont {Deutsch}}, \bibinfo {author} {\bibfnamefont
  {J.}~\bibnamefont {Reichel}}, \ and\ \bibinfo {author} {\bibfnamefont
  {J.~G.~E.}\ \bibnamefont {Harris}},\ }\href {\doibase
  http://dx.doi.org/10.1063/1.4768779} {\bibfield  {journal} {\bibinfo
  {journal} {Appl. Phys. Lett.}\ }\textbf {\bibinfo {volume} {101}},\ \bibinfo
  {eid} {221109} (\bibinfo {year} {2012})}\BibitemShut {NoStop}%
\bibitem [{\citenamefont {Xuereb}\ and\ \citenamefont
  {Domokos}(2012)}]{xuereb_dynamical_2012}%
  \BibitemOpen
  \bibfield  {author} {\bibinfo {author} {\bibfnamefont {A.}~\bibnamefont
  {Xuereb}}\ and\ \bibinfo {author} {\bibfnamefont {P.}~\bibnamefont
  {Domokos}},\ }\href {\doibase 10.1088/1367-2630/14/9/095027} {\bibfield
  {journal} {\bibinfo  {journal} {New J. Phys.}\ }\textbf {\bibinfo {volume}
  {14}},\ \bibinfo {pages} {095027} (\bibinfo {year} {2012})}\BibitemShut
  {NoStop}%
\bibitem [{\citenamefont {Rabl}(2011)}]{Rabl2011}%
  \BibitemOpen
  \bibfield  {author} {\bibinfo {author} {\bibfnamefont {P.}~\bibnamefont
  {Rabl}},\ }\href {\doibase 10.1103/PhysRevLett.107.063601} {\bibfield
  {journal} {\bibinfo  {journal} {Phys. Rev. Lett.}\ }\textbf {\bibinfo
  {volume} {107}},\ \bibinfo {pages} {063601} (\bibinfo {year}
  {2011})}\BibitemShut {NoStop}%
\bibitem [{\citenamefont {K\'om\'ar}\ \emph {et~al.}(2013)\citenamefont
  {K\'om\'ar}, \citenamefont {Bennett}, \citenamefont {Stannigel},
  \citenamefont {Habraken}, \citenamefont {Rabl}, \citenamefont {Zoller},\ and\
  \citenamefont {Lukin}}]{Komar2013}%
  \BibitemOpen
  \bibfield  {author} {\bibinfo {author} {\bibfnamefont {P.}~\bibnamefont
  {K\'om\'ar}}, \bibinfo {author} {\bibfnamefont {S.~D.}\ \bibnamefont
  {Bennett}}, \bibinfo {author} {\bibfnamefont {K.}~\bibnamefont {Stannigel}},
  \bibinfo {author} {\bibfnamefont {S.~J.~M.}\ \bibnamefont {Habraken}},
  \bibinfo {author} {\bibfnamefont {P.}~\bibnamefont {Rabl}}, \bibinfo {author}
  {\bibfnamefont {P.}~\bibnamefont {Zoller}}, \ and\ \bibinfo {author}
  {\bibfnamefont {M.~D.}\ \bibnamefont {Lukin}},\ }\href {\doibase
  10.1103/PhysRevA.87.013839} {\bibfield  {journal} {\bibinfo  {journal} {Phys.
  Rev. A}\ }\textbf {\bibinfo {volume} {87}},\ \bibinfo {pages} {013839}
  (\bibinfo {year} {2013})}\BibitemShut {NoStop}%
\bibitem [{\citenamefont {Ludwig}\ \emph {et~al.}(2012)\citenamefont {Ludwig},
  \citenamefont {Safavi-Naeini}, \citenamefont {Painter},\ and\ \citenamefont
  {Marquardt}}]{Ludwig2012}%
  \BibitemOpen
  \bibfield  {author} {\bibinfo {author} {\bibfnamefont {M.}~\bibnamefont
  {Ludwig}}, \bibinfo {author} {\bibfnamefont {A.~H.}\ \bibnamefont
  {Safavi-Naeini}}, \bibinfo {author} {\bibfnamefont {O.}~\bibnamefont
  {Painter}}, \ and\ \bibinfo {author} {\bibfnamefont {F.}~\bibnamefont
  {Marquardt}},\ }\href {\doibase 10.1103/PhysRevLett.109.063601} {\bibfield
  {journal} {\bibinfo  {journal} {Phys. Rev. Lett.}\ }\textbf {\bibinfo
  {volume} {109}},\ \bibinfo {pages} {063601} (\bibinfo {year}
  {2012})}\BibitemShut {NoStop}%
\bibitem [{\citenamefont {Fan}\ and\ \citenamefont
  {Joannopoulos}(2002)}]{fan_analysis_2002}%
  \BibitemOpen
  \bibfield  {author} {\bibinfo {author} {\bibfnamefont {S.}~\bibnamefont
  {Fan}}\ and\ \bibinfo {author} {\bibfnamefont {J.}~\bibnamefont
  {Joannopoulos}},\ }\href@noop {} {\bibfield  {journal} {\bibinfo  {journal}
  {Phys. Rev. B}\ }\textbf {\bibinfo {volume} {65}} (\bibinfo {year}
  {2002})}\BibitemShut {NoStop}%
\bibitem [{\citenamefont {Antoni}\ \emph {et~al.}(2011)\citenamefont {Antoni},
  \citenamefont {Kuhn}, \citenamefont {Briant}, \citenamefont {Cohadon},
  \citenamefont {Heidmann}, \citenamefont {Braive}, \citenamefont {Beveratos},
  \citenamefont {Abram}, \citenamefont {Gratiet}, \citenamefont {Sagnes},\ and\
  \citenamefont {Robert-Philip}}]{antoni_deformable_2011}%
  \BibitemOpen
  \bibfield  {author} {\bibinfo {author} {\bibfnamefont {T.}~\bibnamefont
  {Antoni}}, \bibinfo {author} {\bibfnamefont {A.~G.}\ \bibnamefont {Kuhn}},
  \bibinfo {author} {\bibfnamefont {T.}~\bibnamefont {Briant}}, \bibinfo
  {author} {\bibfnamefont {P.-F.}\ \bibnamefont {Cohadon}}, \bibinfo {author}
  {\bibfnamefont {A.}~\bibnamefont {Heidmann}}, \bibinfo {author}
  {\bibfnamefont {R.}~\bibnamefont {Braive}}, \bibinfo {author} {\bibfnamefont
  {A.}~\bibnamefont {Beveratos}}, \bibinfo {author} {\bibfnamefont
  {I.}~\bibnamefont {Abram}}, \bibinfo {author} {\bibfnamefont {L.~L.}\
  \bibnamefont {Gratiet}}, \bibinfo {author} {\bibfnamefont {I.}~\bibnamefont
  {Sagnes}}, \ and\ \bibinfo {author} {\bibfnamefont {I.}~\bibnamefont
  {Robert-Philip}},\ }\href {\doibase 10.1364/OL.36.003434} {\bibfield
  {journal} {\bibinfo  {journal} {Opt. Lett.}\ }\textbf {\bibinfo {volume}
  {36}},\ \bibinfo {pages} {3434} (\bibinfo {year} {2011})}\BibitemShut
  {NoStop}%
\bibitem [{\citenamefont {Antoni}\ \emph {et~al.}(2012)\citenamefont {Antoni},
  \citenamefont {Makles}, \citenamefont {Braive}, \citenamefont {Briant},
  \citenamefont {Cohadon}, \citenamefont {Sagnes}, \citenamefont
  {Robert-Philip},\ and\ \citenamefont {Heidmann}}]{antoni_nonlinear_2012}%
  \BibitemOpen
  \bibfield  {author} {\bibinfo {author} {\bibfnamefont {T.}~\bibnamefont
  {Antoni}}, \bibinfo {author} {\bibfnamefont {K.}~\bibnamefont {Makles}},
  \bibinfo {author} {\bibfnamefont {R.}~\bibnamefont {Braive}}, \bibinfo
  {author} {\bibfnamefont {T.}~\bibnamefont {Briant}}, \bibinfo {author}
  {\bibfnamefont {P.-F.}\ \bibnamefont {Cohadon}}, \bibinfo {author}
  {\bibfnamefont {I.}~\bibnamefont {Sagnes}}, \bibinfo {author} {\bibfnamefont
  {I.}~\bibnamefont {Robert-Philip}}, \ and\ \bibinfo {author} {\bibfnamefont
  {A.}~\bibnamefont {Heidmann}},\ }\href
  {http://stacks.iop.org/0295-5075/100/i=6/a=68005} {\bibfield  {journal}
  {\bibinfo  {journal} {EPL (Europhysics Letters)}\ }\textbf {\bibinfo {volume}
  {100}},\ \bibinfo {pages} {68005} (\bibinfo {year} {2012})}\BibitemShut
  {NoStop}%
\bibitem [{\citenamefont {Bui}\ \emph {et~al.}(2012)\citenamefont {Bui},
  \citenamefont {Zheng}, \citenamefont {Hoch}, \citenamefont {Lee},
  \citenamefont {Harris},\ and\ \citenamefont {Wei~Wong}}]{BuiPhotXtalAPL2012}%
  \BibitemOpen
  \bibfield  {author} {\bibinfo {author} {\bibfnamefont {C.~H.}\ \bibnamefont
  {Bui}}, \bibinfo {author} {\bibfnamefont {J.}~\bibnamefont {Zheng}}, \bibinfo
  {author} {\bibfnamefont {S.~W.}\ \bibnamefont {Hoch}}, \bibinfo {author}
  {\bibfnamefont {L.~Y.~T.}\ \bibnamefont {Lee}}, \bibinfo {author}
  {\bibfnamefont {J.~G.~E.}\ \bibnamefont {Harris}}, \ and\ \bibinfo {author}
  {\bibfnamefont {C.}~\bibnamefont {Wei~Wong}},\ }\href {\doibase
  http://dx.doi.org/10.1063/1.3658731} {\bibfield  {journal} {\bibinfo
  {journal} {Appl. Phys. Lett.}\ }\textbf {\bibinfo {volume} {100}},\ \bibinfo
  {eid} {021110} (\bibinfo {year} {2012})}\BibitemShut {NoStop}%
\bibitem [{\citenamefont {Huang}\ \emph {et~al.}(2007)\citenamefont {Huang},
  \citenamefont {Zhou},\ and\ \citenamefont
  {Chang-Hasnain}}]{huang_surface-emitting_2007}%
  \BibitemOpen
  \bibfield  {author} {\bibinfo {author} {\bibfnamefont {M.~C.}\ \bibnamefont
  {Huang}}, \bibinfo {author} {\bibfnamefont {Y.}~\bibnamefont {Zhou}}, \ and\
  \bibinfo {author} {\bibfnamefont {C.~J.}\ \bibnamefont {Chang-Hasnain}},\
  }\href {\doibase 10.1038/nphoton.2006.80} {\bibfield  {journal} {\bibinfo
  {journal} {Nat Photon}\ }\textbf {\bibinfo {volume} {1}},\ \bibinfo {pages}
  {119} (\bibinfo {year} {2007})}\BibitemShut {NoStop}%
\bibitem [{\citenamefont {Brückner}\ \emph {et~al.}(2010)\citenamefont
  {Brückner}, \citenamefont {Friedrich}, \citenamefont {Clausnitzer},
  \citenamefont {Britzger}, \citenamefont {Burmeister}, \citenamefont
  {Danzmann}, \citenamefont {Kley}, \citenamefont {Tünnermann},\ and\
  \citenamefont {Schnabel}}]{bruckner_realization_2010}%
  \BibitemOpen
  \bibfield  {author} {\bibinfo {author} {\bibfnamefont {F.}~\bibnamefont
  {Brückner}}, \bibinfo {author} {\bibfnamefont {D.}~\bibnamefont {Friedrich}},
  \bibinfo {author} {\bibfnamefont {T.}~\bibnamefont {Clausnitzer}}, \bibinfo
  {author} {\bibfnamefont {M.}~\bibnamefont {Britzger}}, \bibinfo {author}
  {\bibfnamefont {O.}~\bibnamefont {Burmeister}}, \bibinfo {author}
  {\bibfnamefont {K.}~\bibnamefont {Danzmann}}, \bibinfo {author}
  {\bibfnamefont {E.-B.}\ \bibnamefont {Kley}}, \bibinfo {author}
  {\bibfnamefont {A.}~\bibnamefont {Tünnermann}}, \ and\ \bibinfo {author}
  {\bibfnamefont {R.}~\bibnamefont {Schnabel}},\ }\href {\doibase
  10.1103/PhysRevLett.104.163903} {\bibfield  {journal} {\bibinfo  {journal}
  {Phys. Rev. Lett.}\ }\textbf {\bibinfo {volume} {104}} (\bibinfo {year}
  {2010}),\ 10.1103/PhysRevLett.104.163903}\BibitemShut {NoStop}%
\bibitem [{\citenamefont {Lu}\ \emph {et~al.}(2010)\citenamefont {Lu},
  \citenamefont {Sedgwick}, \citenamefont {Karagodsky}, \citenamefont {Chase},\
  and\ \citenamefont {Chang-Hasnain}}]{lu_planar_2010}%
  \BibitemOpen
  \bibfield  {author} {\bibinfo {author} {\bibfnamefont {F.}~\bibnamefont
  {Lu}}, \bibinfo {author} {\bibfnamefont {F.~G.}\ \bibnamefont {Sedgwick}},
  \bibinfo {author} {\bibfnamefont {V.}~\bibnamefont {Karagodsky}}, \bibinfo
  {author} {\bibfnamefont {C.}~\bibnamefont {Chase}}, \ and\ \bibinfo {author}
  {\bibfnamefont {C.~J.}\ \bibnamefont {Chang-Hasnain}},\ }\href
  {http://www.opticsinfobase.org/abstract.cfm?uri=oe-18-12-12606} {\bibfield
  {journal} {\bibinfo  {journal} {Opt. Express}\ }\textbf {\bibinfo {volume}
  {18}},\ \bibinfo {pages} {12606} (\bibinfo {year} {2010})}\BibitemShut
  {NoStop}%
\bibitem [{\citenamefont {Fattal}\ \emph {et~al.}(2010)\citenamefont {Fattal},
  \citenamefont {Li}, \citenamefont {Peng}, \citenamefont {Fiorentino},\ and\
  \citenamefont {Beausoleil}}]{fattal_flat_2010}%
  \BibitemOpen
  \bibfield  {author} {\bibinfo {author} {\bibfnamefont {D.}~\bibnamefont
  {Fattal}}, \bibinfo {author} {\bibfnamefont {J.}~\bibnamefont {Li}}, \bibinfo
  {author} {\bibfnamefont {Z.}~\bibnamefont {Peng}}, \bibinfo {author}
  {\bibfnamefont {M.}~\bibnamefont {Fiorentino}}, \ and\ \bibinfo {author}
  {\bibfnamefont {R.~G.}\ \bibnamefont {Beausoleil}},\ }\href {\doibase
  10.1038/nphoton.2010.116} {\bibfield  {journal} {\bibinfo  {journal} {Nat.
  Photonics}\ }\textbf {\bibinfo {volume} {4}},\ \bibinfo {pages} {466}
  (\bibinfo {year} {2010})}\BibitemShut {NoStop}%
\bibitem [{\citenamefont
  {Chang-Hasnain}(2011)}]{chang-hasnain_high-contrast_2011}%
  \BibitemOpen
  \bibfield  {author} {\bibinfo {author} {\bibfnamefont {C.~J.}\ \bibnamefont
  {Chang-Hasnain}},\ }\href@noop {} {\bibfield  {journal} {\bibinfo  {journal}
  {Semicond. Sci. Technol.}\ }\textbf {\bibinfo {volume} {26}} (\bibinfo {year}
  {2011})},\ \bibinfo {note} {014043}\BibitemShut {NoStop}%
\bibitem [{\citenamefont {Kemiktarak}\ \emph
  {et~al.}(2012{\natexlab{a}})\citenamefont {Kemiktarak}, \citenamefont
  {Metcalfe}, \citenamefont {Durand},\ and\ \citenamefont
  {Lawall}}]{Kemiktarak2012}%
  \BibitemOpen
  \bibfield  {author} {\bibinfo {author} {\bibfnamefont {U.}~\bibnamefont
  {Kemiktarak}}, \bibinfo {author} {\bibfnamefont {M.}~\bibnamefont
  {Metcalfe}}, \bibinfo {author} {\bibfnamefont {M.}~\bibnamefont {Durand}}, \
  and\ \bibinfo {author} {\bibfnamefont {J.}~\bibnamefont {Lawall}},\ }\href
  {\doibase 10.1063/1.3684248} {\bibfield  {journal} {\bibinfo  {journal}
  {Appl. Phys. Lett.}\ }\textbf {\bibinfo {volume} {100}},\ \bibinfo {pages}
  {061124} (\bibinfo {year} {2012}{\natexlab{a}})}\BibitemShut {NoStop}%
\bibitem [{\citenamefont {Karagodsky}\ \emph {et~al.}(2010)\citenamefont
  {Karagodsky}, \citenamefont {Sedgwick},\ and\ \citenamefont
  {Chang-Hasnain}}]{karagodsky_theoretical_2010}%
  \BibitemOpen
  \bibfield  {author} {\bibinfo {author} {\bibfnamefont {V.}~\bibnamefont
  {Karagodsky}}, \bibinfo {author} {\bibfnamefont {F.~G.}\ \bibnamefont
  {Sedgwick}}, \ and\ \bibinfo {author} {\bibfnamefont {C.~J.}\ \bibnamefont
  {Chang-Hasnain}},\ }\href {\doibase 10.1364/OE.18.016973} {\bibfield
  {journal} {\bibinfo  {journal} {Opt. Express}\ }\textbf {\bibinfo {volume}
  {18}},\ \bibinfo {pages} {16973} (\bibinfo {year} {2010})}\BibitemShut
  {NoStop}%
\bibitem [{\citenamefont {Moharam}\ and\ \citenamefont
  {Gaylord}(1981)}]{moharam_rigorous_1981}%
  \BibitemOpen
  \bibfield  {author} {\bibinfo {author} {\bibfnamefont {M.}~\bibnamefont
  {Moharam}}\ and\ \bibinfo {author} {\bibfnamefont {T.}~\bibnamefont
  {Gaylord}},\ }\href@noop {} {\bibfield  {journal} {\bibinfo  {journal}
  {{JOSA}}\ }\textbf {\bibinfo {volume} {71}},\ \bibinfo {pages} {811?818}
  (\bibinfo {year} {1981})}\BibitemShut {NoStop}%
\bibitem [{\citenamefont {Stambaugh}\ \emph {et~al.}(2014)\citenamefont
  {Stambaugh}, \citenamefont {Durand}, \citenamefont {Kemiktarak},\ and\
  \citenamefont {Lawall}}]{stambaugh2014}%
  \BibitemOpen
  \bibfield  {author} {\bibinfo {author} {\bibfnamefont {C.}~\bibnamefont
  {Stambaugh}}, \bibinfo {author} {\bibfnamefont {M.}~\bibnamefont {Durand}},
  \bibinfo {author} {\bibfnamefont {U.}~\bibnamefont {Kemiktarak}}, \ and\
  \bibinfo {author} {\bibfnamefont {J.}~\bibnamefont {Lawall}},\ }\href@noop {}
  {\  (\bibinfo {year} {2014})},\ \bibinfo {note} {(submitted for
  publication)}\BibitemShut {NoStop}%
\bibitem [{\citenamefont {Kemiktarak}\ \emph
  {et~al.}(2012{\natexlab{b}})\citenamefont {Kemiktarak}, \citenamefont
  {Durand}, \citenamefont {Metcalfe},\ and\ \citenamefont
  {Lawall}}]{Kemiktarak2012a}%
  \BibitemOpen
  \bibfield  {author} {\bibinfo {author} {\bibfnamefont {U.}~\bibnamefont
  {Kemiktarak}}, \bibinfo {author} {\bibfnamefont {M.}~\bibnamefont {Durand}},
  \bibinfo {author} {\bibfnamefont {M.}~\bibnamefont {Metcalfe}}, \ and\
  \bibinfo {author} {\bibfnamefont {J.}~\bibnamefont {Lawall}},\ }\href
  {\doibase 10.1088/1367-2630/14/12/125010} {\bibfield  {journal} {\bibinfo
  {journal} {New J. Phys.}\ }\textbf {\bibinfo {volume} {14}},\ \bibinfo
  {pages} {125010} (\bibinfo {year} {2012}{\natexlab{b}})}\BibitemShut
  {NoStop}%
\bibitem [{\citenamefont {Fowles}(1975)}]{Fowles}%
  \BibitemOpen
  \bibfield  {author} {\bibinfo {author} {\bibfnamefont {G.}~\bibnamefont
  {Fowles}},\ }\href {http://books.google.com/books?id=SL1n9TuJ5YMC} {\emph
  {\bibinfo {title} {Introduction to Modern Optics}}},\ Dover Books on Physics
  Series\ (\bibinfo  {publisher} {Dover Publications},\ \bibinfo {year}
  {1975})\BibitemShut {NoStop}%
\bibitem [{\citenamefont {Spencer}\ and\ \citenamefont
  {Lamb}(1972)}]{Spencer_Lamb_PRA_1972}%
  \BibitemOpen
  \bibfield  {author} {\bibinfo {author} {\bibfnamefont {M.~B.}\ \bibnamefont
  {Spencer}}\ and\ \bibinfo {author} {\bibfnamefont {W.~E.}\ \bibnamefont
  {Lamb}},\ }\href@noop {} {\bibfield  {journal} {\bibinfo  {journal} {Phys.
  Rev. A}\ }\textbf {\bibinfo {volume} {5}},\ \bibinfo {pages} {893} (\bibinfo
  {year} {1972})}\BibitemShut {NoStop}%
\bibitem [{\citenamefont {Fader}(1985)}]{fader_theory_1985}%
  \BibitemOpen
  \bibfield  {author} {\bibinfo {author} {\bibfnamefont {W.}~\bibnamefont
  {Fader}},\ }\href
  {http://ieeexplore.ieee.org/xpls/abs_all.jsp?arnumber=1072577} {\bibfield
  {journal} {\bibinfo  {journal} {Quantum Electronics, {IEEE} Journal of}\
  }\textbf {\bibinfo {volume} {21}},\ \bibinfo {pages} {1838} (\bibinfo {year}
  {1985})}\BibitemShut {NoStop}%
\bibitem [{\citenamefont {Chow}(1986)}]{chow_composite-resonator_1986}%
  \BibitemOpen
  \bibfield  {author} {\bibinfo {author} {\bibfnamefont {W.~W.}\ \bibnamefont
  {Chow}},\ }\href
  {http://ieeexplore.ieee.org/xpls/abs_all.jsp?arnumber=1073104} {\bibfield
  {journal} {\bibinfo  {journal} {Quantum Electronics, {IEEE} Journal of}\
  }\textbf {\bibinfo {volume} {22}},\ \bibinfo {pages} {1174} (\bibinfo {year}
  {1986})}\BibitemShut {NoStop}%
\bibitem [{\citenamefont {Kogelnik}\ and\ \citenamefont
  {Li}(1966)}]{kogelnik_laser_1966}%
  \BibitemOpen
  \bibfield  {author} {\bibinfo {author} {\bibfnamefont {H.}~\bibnamefont
  {Kogelnik}}\ and\ \bibinfo {author} {\bibfnamefont {T.}~\bibnamefont {Li}},\
  }\href@noop {} {\bibfield  {journal} {\bibinfo  {journal} {Proc. IEEE}\
  }\textbf {\bibinfo {volume} {54}},\ \bibinfo {pages} {1312} (\bibinfo {year}
  {1966})}\BibitemShut {NoStop}%
\bibitem [{\citenamefont {Siegman}(1986)}]{Siegman1986}%
  \BibitemOpen
  \bibfield  {author} {\bibinfo {author} {\bibfnamefont {A.}~\bibnamefont
  {Siegman}},\ }\href@noop {} {\emph {\bibinfo {title} {Lasers}}}\ (\bibinfo
  {publisher} {University Science Books, Sausalito},\ \bibinfo {year}
  {1986})\BibitemShut {NoStop}%
\bibitem [{\citenamefont {Wilson}\ \emph {et~al.}(2009)\citenamefont {Wilson},
  \citenamefont {Regal}, \citenamefont {Papp},\ and\ \citenamefont
  {Kimble}}]{Wilson2009}%
  \BibitemOpen
  \bibfield  {author} {\bibinfo {author} {\bibfnamefont {D.}~\bibnamefont
  {Wilson}}, \bibinfo {author} {\bibfnamefont {C.}~\bibnamefont {Regal}},
  \bibinfo {author} {\bibfnamefont {S.}~\bibnamefont {Papp}}, \ and\ \bibinfo
  {author} {\bibfnamefont {H.}~\bibnamefont {Kimble}},\ }\href {\doibase
  10.1103/PhysRevLett.103.207204} {\bibfield  {journal} {\bibinfo  {journal}
  {Phys. Rev. Lett.}\ }\textbf {\bibinfo {volume} {103}} (\bibinfo {year}
  {2009}),\ 10.1103/PhysRevLett.103.207204}\BibitemShut {NoStop}%
\end{thebibliography}%

\end{document}